\documentclass[aps,pra,twocolumn,amsfonts,amsmath,showpacs,floatfix,nofootinbib,groupedaddress,superscriptaddress]{revtex4}
\usepackage{amssymb,graphics,graphicx,times,bm,bbm,multirow,color}

\begin{document}

\title{Quantum Process Tomography: Resource Analysis of Different Strategies}
\author{M. Mohseni}
\affiliation{Department of Chemistry and Chemical Biology, Harvard
University, 12 Oxford St., Cambridge, MA 02138, USA}
\affiliation{Department of Chemistry, University of Southern
California, Los Angeles, CA 90089, USA}
\author{A. T. Rezakhani}
\affiliation{Center for Quantum Information Science and Technology,
and Departments of Chemistry and Physics, University of Southern
California, Los Angeles, CA 90089, USA}
\affiliation{Institute for
Quantum Information Science, University of Calgary, Alberta T2N 1N4,
Canada}
\author{D. A. Lidar}
\affiliation{Department of Chemistry, University of Southern
California, Los Angeles, CA 90089, USA}
\affiliation{Departments of Physics and Electrical Engineering, University of Southern California, Los
Angeles, CA 90089, USA}

\begin{abstract}
Characterization of quantum dynamics is a fundamental problem in
quantum physics and quantum information science. Several methods are
known which achieve this goal, namely Standard Quantum Process
Tomography (SQPT), Ancilla-Assisted Process Tomography (AAPT), and
the recently proposed scheme of Direct Characterization of Quantum
Dynamics (DCQD). Here, we review these schemes and analyze them with
respect to some of the physical resources they require. Although a
reliable figure-of-merit for process characterization is not yet
available, our analysis can provide a benchmark which is necessary
for choosing the scheme that is the most appropriate in a given
situation, with given resources. As a result, we conclude that for
quantum systems where two-body interactions are not naturally
available, SQPT is the most efficient scheme. However, for quantum
systems with controllable two-body interactions, the DCQD scheme is
more efficient than other known QPT schemes in terms of the total
number of required elementary quantum operations.
\end{abstract}

\maketitle

\section{Introduction}

\label{intro}

Characterization of quantum dynamical systems is a central task in
quantum control and quantum information processing. Knowledge of the
\emph{state} of a quantum system is indispensable in
identification/verification of experimental outcomes. Quantum state
tomography has been developed as a general scheme to accomplish this
task \cite{nielsen-book}. In this method an arbitrary and unknown
quantum state can be estimated by measuring the expectation values
of a set of observables on an ensemble of identical quantum systems
prepared in the same initial state. Identification of an unknown
quantum \emph{process} acting on a quantum system is another vital
task in coherent control of the dynamics. This task is especially
crucial in verifying the performance of a quantum device in the
presence of decoherence. In general, procedures for characterization
of quantum dynamical maps are known as quantum process tomography
(QPT)---for a review of quantum tomography see
Refs.~\cite{d'ariano-qt,d'ariano-cqd,artiles}.

There are two types of methods for characterization of quantum dynamics:
direct and indirect. In indirect methods, information about the underlying
quantum process is mapped onto the state of some probe quantum system(s),
and the process is reconstructed via quantum state tomography on the output
states. We call these methods indirect since they \emph{require} quantum
state tomography in order to reconstruct a quantum process. A further
unavoidable step in indirect methods is the application of an inversion map
on the final output data. Standard Quantum Process Tomography (SQPT) \cite%
{nielsen-book,chuang-sqpt,poyatos-sqpt} and Ancilla-Assisted Process
Tomography (AAPT)
\cite{leung,d'ariano-aapt,altepeter-aapt,d'ariano-faithful} belong
to this class. On the other hand, in direct methods each
experimental outcome directly provides information about properties
of the underlying dynamics, without the need for state tomography.
In the last decade, there has been a growing interest in the
development of such direct methods for obtaining specific
information about the states and dynamics of quantum systems, such
as estimation of general functions of a quantum state
\cite{ekert-direct}, detection of quantum entanglement
\cite{horodecki-direct}, measurement of nonlinear properties of
bipartite quantum states \cite{bovino-direct}, estimation of the
average fidelity of a quantum gate or process
\cite{Emerson-direct,Hofmann-direct}, and universal source coding
and data compression \cite{bennett-compression}. The method of
Direct Characterization of Quantum Dynamics (DCQD) \cite
{mohseni-dcqd1,mohseni-dcqd2,MasoudThesis,WangExDCQD07} is the first
scheme which provides a full characterization of (closed or open)
quantum systems without performing any state tomography. In this
method each probe system and the corresponding measurements are
devised in such a way that the final probability distributions of
the outcomes become more directly related to specific classes of the
elements of the dynamics. A complete set of probe states can then be
utilized to fully characterize the unknown quantum dynamical map.
The preparation of the probe systems and the measurement schemes are
based on quantum error-detection techniques. By construction, this
error-detection based measurement allows for direct estimation of
quantum dynamics such that the need for a complete inversion of
final results does not arise. 
Moreover, by construction, DCQD can be efficiently applied to
partial characterization of quantum dynamics. For example, as
demonstrated in Refs.~\cite{MasoudThesis,mohseni-rezakhani-aspuru},
the DCQD scheme can be used for Hamiltonian identification, and also
for simultaneous determination of the relaxation time $T_{1}$ and
the dephasing time $T_{2}$ in two-level systems. A
proof-of-principle optical realization of DCQD via a Hong-Ou-Mandel
interferometer has also been reported \cite{WangExDCQD07}. Recently,
direct approaches for efficient partial/selective estimation of
quantum processes based on random sampling have been introduced
\cite{Emerson07}. Application of the direct QPT methods to the
efficient parameter estimation of many-body quantum Hamiltonian
systems is also of special interest for practical purposes, and will
be addressed in another publication
\cite{mohseni-rezakhani-aspuru08}.

In this work, we review all known methods for \emph{complete}
characterization of quantum dynamics, and analyze the required
physical resources that arise in preparation and quantum
measurements. To the best of our knowledge, this is the first
complexity analysis of different QPT schemes. We conclude that, for
quantum systems with controllable single- and two-body interactions,
the DCQD scheme is more efficient than the other known QPT schemes,
in the sense that it requires a smaller total number of experimental
configurations and/or elementary quantum operations. However, for
quantum systems where two-body interactions are not naturally
available (e.g., photons), the DCQD scheme and (non-separable) AAPT
cannot be implemented or simulated with high efficiency, and the
SQPT scheme is in this case the most efficient.

The structure of this paper is as follows. In Sec.~\ref{QDM-sec}, we briefly
review the concept of a quantum dynamical map. In the subsequent sections, %
\ref{SQPT-sec}, \ref{AAPT-sec} and \ref{DCQD-sec}, we provide a review of
the SQPT, AAPT, and DCQD schemes, respectively. Since SQPT has been
extensively described in earlier literature, we provide more detail about
the AAPT and DCQD schemes. Specifically, we provide a comprehensive
discussion of the different alternative AAPT measurement strategies, i.e.,
those utilizing either joint separable measurements, mutually unbiased bases
measurements, or generalized measurements. For simplicity, we assume that
all quantum operations, including preparations and measurements, are ideal;
i.e., we do not consider the effect of decoherence during the implementation
of a QPT scheme. In the final section of the paper --- Sec.~\ref%
{discussion-sec} --- we present a detailed discussion and comparison
of the different QPT strategies.


\section{Quantum dynamical maps}

\label{QDM-sec}

Under rather general conditions (but assuming a factorized initial
system-bath state) the dynamics of an open quantum system can be described
by a completely-positive linear map, as follows:
\begin{equation}
\mathcal{E}(\rho )=\sum_{i}A_{i}\rho A_{i}^{\dagger },
\end{equation}%
where $\rho $ is the initial state of the system [$\rho \in \mathcal{B}(%
\mathcal{H})$, the space of linear operators acting on $\mathcal{H}$] and $%
\sum_{i}A_{i}^{\dagger }A_{i}\leqslant I$ guarantees that Tr$\mathcal{E}(\rho )\leqslant 1$ \cite%
{nielsen-book}. Suppose that $\{E_{i}\}_{i=0}^{d^{2}-1}$ is a set of
fixed Hermitian basis operators for $\mathcal{B}(\mathcal{H})$,
which satisfy the orthogonality condition
\begin{equation}
\text{Tr}(E_{i}^{\dagger }E_{j})=d\delta _{ij}.
\end{equation}%
For example, for a multi-qubit system the $E_{i}$'s can be tensor products
of identity and Pauli matrices. The $A_{i}$ operators can be decomposed as $%
A_{i}=\sum_{m}a_{im}E_{m}$, and therefore we have
\begin{equation}
\mathcal{E}(\rho )=\sum_{mn=0}^{d^{2}-1}\chi _{mn}E_{m}\rho E_{n}^{\dagger },
\end{equation}%
where $\chi _{mn}=\sum_{ij}a_{mi}a_{nj}^{\ast }$. The positive superoperator
$\bm{\chi}$ encompasses all the information about the map $\mathcal{E}$ with
respect to the $\{E_{i}\}$ basis, i.e., characterization of $\mathcal{E}$ is
equivalent to a determination of the $d^{4}$ independent matrix elements of $%
\bm{\chi}$, where the $E_{i}$ play the role of observables. When the map $%
\mathcal{E}$ is trace-preserving, i.e., $\sum_{i}A_{i}^{\dagger }A_{i}=I$,
the corresponding superoperator $\bm{\chi}$ has only $d^{4}-d^{2}$
independent elements. Hereafter, we restrict our attention only to the $n$%
-qubit case, i.e., $d=2^{n}$.


\section{Standard Quantum Process Tomography}

\label{SQPT-sec}

\begin{figure}[tp]
\includegraphics[width=5.2cm,height=.6cm]{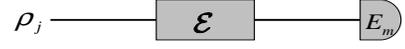}
\caption{Schematic of SQPT. An ensemble of states $\{\protect\rho_j\}$ are
prepared and each of them is subjected to the map $\mathcal{E}$, and then to
the measurements $\{E_m\}$.}
\label{sqpt-f}
\end{figure}
The central idea of SQPT is to prepare $d^{2}$ linearly-independent inputs $%
\{\rho _{k}\}_{k=0}^{d^{2}-1}$ and then measure the output states $\mathcal{E%
}(\rho _{k})$ by using quantum state tomography \cite%
{nielsen-book,chuang-sqpt,poyatos-sqpt}. SQPT has been experimentally
demonstrated in liquid-state NMR \cite%
{childs-nmr,boulant-lindblad,weinstein-qpt}, optical \cite%
{steinberg-bell,obrien-cnot}, atomic \cite{steinberg-qpt2} and solid-state
systems \cite{howard}. Since the map $\mathcal{E}$ is linear, it can in
principle be reconstructed from the measured data by a proper inversion. Let
$\{\rho _{k}\}_{k=0}^{d^{2}-1}$ be a linearly independent basis set of
operators for the space of $d\times d$ linear operators. A convenient choice
is $\rho _{k}=|m\rangle \langle n|$, where $\{|m\rangle \}_{m=0}^{d-1}$ is
an orthonormal basis for $\mathcal{H}$. The coherence $|m\rangle \langle n|$
can be reconstructed from four populations:\ $|m\rangle \langle n|=|+\rangle
\langle +|+|-\rangle \langle -|-[|m\rangle \langle m|+|n\rangle \langle
n|](1+i)/2$, where $|+\rangle =(|m\rangle +|n\rangle )/\sqrt{2}$ and $%
|-\rangle =(|m\rangle +i|n\rangle )/\sqrt{2}$. Linearity of $\mathcal{E}$
then implies that measurement of $\mathcal{E}(|+\rangle \langle +|)$, $%
\mathcal{E}(|-\rangle \langle -|)$, $\mathcal{E}(|m\rangle \langle m|)$, and
$\mathcal{E}(|n\rangle \langle n|)$ suffices for the determination of $%
\mathcal{E}(|m\rangle \langle n|)$. In addition, every $\mathcal{E}(\rho
_{k})$ can be expressed in terms of a linear combination of basis states, as
$\mathcal{E}(\rho _{k})=\sum_{l}\lambda _{kl}\rho _{l}$. The parameters $%
\lambda _{kl}$ contain the measurement results, and can be understood as the
expectation values of the fixed-basis operators $E_{k}\text{: }$%
\begin{equation}
\lambda _{kl}=\text{Tr}(E_{k}\mathcal{E}(\rho _{l})),  \label{eq:lambda_kl}
\end{equation}
when $E_{k}=\rho _{k}$. This choice of the $E_{k}$ is natural, since the $%
\rho _{k}$ are Hermitian operators and thus they are valid
observables. If we combine this with the relation $E_{m}\rho
_{k}E_{n}^{\dagger
}=\sum_{l}B_{mn,lk}\rho _{l}$, the following equation can be obtained: $%
\sum_{mn}B_{mn,lk}\chi _{mn}=\lambda _{kl}$. This in turn can be written in
the following matrix form:
\begin{equation}
\bm{B}\bm{\chi}=\bm{\lambda},  \label{sqpt}
\end{equation}%
where the $d^{4}\times d^{4}$-dimensional matrix $\bm{B}$ is determined by
the choice of bases $\{\rho _{k}\}$ and $\{E_{m}\}$, and the $d^{4}$%
-dimensional vector $\bm{\lambda}$ is determined from the state tomography
experiments. The superoperator $\bm{\chi}$ can thus be determined by
inversion of Eq.~(\ref{sqpt}), but in general $\bm{\chi}$ is not uniquely
determined by this equation.{}

Figure~\ref{sqpt-f} illustrates the SQPT scheme. Let us determine the
resources this scheme requires. In general, SQPT involves preparation of $%
d^{2}$ linearly independent inputs $\{\rho _{l}\}$, each of which is
subjected to the quantum process $\mathcal{E}$, followed by quantum state
tomography on the corresponding outputs. As we saw above, for each $\rho
_{l} $ we must measure the expectation values of the $d^{2}$ fixed-basis
operators $\{E_{k}\}$ in the state $\mathcal{E}(\rho _{l})$. Thus the total
number of required measurements amounts to $d^{4}$. Since measurement of an
expectation value cannot be done on a single copy of a system, throughout
this paper, whenever we use the term \textquotedblleft
measurement\textquotedblright\ we implicitly mean measurement on an \emph{%
ensemble} of identically prepared quantum systems corresponding to a
given experimental setting.


\section{Ancilla-Assisted Process Tomography}

\label{AAPT-sec}

In principle, there is an intrinsic analogy between quantum
\textit{state} tomography schemes and QPT. This analogy is based
upon the well-known Choi-Jamiolkowski isomorphism
\cite{choi-jamilolkowksi}, which establishes a correspondence
between completely-positive quantum maps (or operations) and quantum
states, $\mathcal{E}\rightarrow \rho_{\mathcal{E}}$, as follows:
\begin{eqnarray}
\rho_{\mathcal{E}}\equiv (\mathcal{E}\otimes I)(|\Phi^+\rangle\langle\Phi^+|),
\end{eqnarray}
where $|\Phi^+\rangle=\sum_{i=1}^d
\frac{1}{\sqrt{d}}|i\rangle\otimes |i\rangle$ is the maximally
entangled state of the system and an ancilla with the same size.
This one-to-one map enables all of the theorems about quantum
operations directly to be derived from those of quantum states
\cite{arrighi}. In this way, one can consider a quantum process as a
quantum state (in a larger Hilbert space). \footnote{Choosing
$\{E_m\}=\{|i\rangle\langle j|\}$ results in:
$\bm{\chi}=d\rho_{\mathcal{E}}$ \cite{gilchrist}.} Therefore, the
identification of the original map $\mathcal{E}$ is equivalent to
the characterization of the corresponding state
$\rho_{\mathcal{E}}$. In other words, the problem of quantum process
tomography can naturally be reduced to the problem of quantum state
tomography, and hence, all state identification techniques can be
applied to the characterization of quantum processes as well. The
AAPT scheme was built exactly upon this basis.

Generally, within the AAPT scheme, we attach an auxiliary system
(ancilla), $B$, to our principal system, $A$, and prepare the
combined system in a single state such that complete information
about the dynamics can be imprinted on the final state
\cite{d'ariano-aapt,altepeter-aapt}. Then by performing quantum
state tomography in the extended Hilbert space of
$\mathcal{H}_{AB}$, one can extract complete information about the
unknown map acting on the principal system. In principle, the input
state of the system and ancilla can be prepared in either an
entangled mixed state (entanglement-assisted) or a separable mixed
state. Intuitively, the input state in AAPT must be faithful enough
to the map $\mathcal{E}$ such that by quantum state tomography on
the outputs one can identify $\mathcal{E}$ completely and
unambiguously \cite{d'ariano-faithful}. This faithfulness condition
can formalized. Indeed, it is easy to show that a state $\rho $ can
be used as input for AAPT iff $\rho $ has maximal Schmidt number,
i.e. $\text{Sch}(\rho )=d^{2}$ \cite{altepeter-aapt}. \footnote{Any
operator $Q$ acting on a
 bipartite system $AB$ can be decomposed as $%
Q=\sum_{l=1}^{\text{Sch}(Q)}s_{l}A_{l}\otimes B_{l}$, where the
$s_{l}$ are all non-negative numbers, and $\{A_{l}\}$ and
$\{B_{l}\}$ are orthonormal operator bases for the systems $A$ and
$B$, respectively \cite{nielsen-dynamics}. $\text{Sch}(Q)$ is
defined as the number of terms in the Schmidt decomposition of $Q$.}
The faithfulness condition is nothing but an
invertibility condition. That is, because of linearity of the map $\mathcal{E%
}\otimes I,$ the information is imprinted on the elements of the final
output states linearly. By performing state tomography on the output state $(%
\mathcal{E}\otimes I)(\rho )$, we obtain a set of linear relations among
possible measurement outcomes and the elements of $\mathcal{E}$ and $\rho $.
The matrix $\rho $ must be chosen such that an inversion becomes possible;
thus one can solve the set of linear equations for $\mathcal{E}$ \cite%
{d'ariano-aapt}. We provide more details below.

It should be noted that the faithfulness condition is different from
entanglement. In fact, almost all states of the combined system $AB$
(excluding product states) may be used for AAPT, because the set of
states with Schmidt number less than $d^{2}$ is of zero measure.
This means that entanglement is not a necessary property of the
input state $\rho $ in AAPT. Indeed, many of the viable input states
are not entangled, such as Werner states \cite{altepeter-aapt}.
However, it has been argued and also experimentally verified that
use of maximally entangled pure states offers the best performance.
That is, even though in principle any faithful state can be used in
AAPT, the propagation of experimental errors from the
measurement outcomes to actual estimation of $(\mathcal{E}\otimes I)(\rho )$%
, due to the inversion process, dictates that different faithful input
states can produce very different errors.

In order to develop a good faithfulness measure, one can consider a general
property of a typical input state $\rho $, such that the output states
generated by different quantum dynamical processes have maximum distance;
e.g., different output states should be nearly orthogonal. More
specifically, we note that the experimental error amplification is related
to the inversion, which in turn depends on the multiplication by the inverse
of the eigenvalues of $\rho $, $s_{l}^{-1}$. Then, the smaller the
eigenvalues the higher the amplification of experimental errors. This fact
has led to the following definition:
\[
F(\rho )=\text{Tr}(\rho ^{2})=\sum_{l=1}^{d^{2}}s_{l}^{2},
\]%
as a proper measure of faithfulness \cite{d'ariano-aapt}. This, indeed, is
exactly the purity of the state $\rho $. As a consequence, this implies that
the optimal (in the sense of minimal experimental errors, as explained
above) faithful input states are pure states with maximal Schmidt number and
$s_{l}=1/\sqrt{d}$, i.e., maximally entangled pure states.
\begin{figure}[tp]
\includegraphics[width=7.2cm,height=2.1cm]{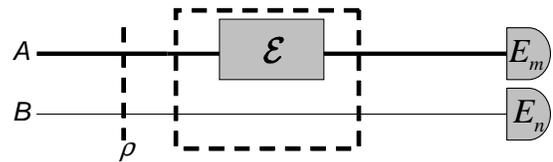}
\caption{Schematic diagram of separable AAPT. An ensemble of systems is
prepared in the same quantum state $\protect\rho $. Next, they are subjected
to the map $\mathcal{E}\otimes I$. Finally the operators $\{E_{j}\}$ are
measured on both the system and the ancilla, which results in the required
joint probability distributions or expectation values.}
\label{aapt-f}
\end{figure}

The Hilbert space of the input state in AAPT is $\mathcal{H}_{AB}\equiv
\mathcal{H}_{A}{\otimes }\mathcal{H}_{B}$. At the output, one can realize
the required quantum state tomography by either separable measurements
(separable AAPT), i.e., joint measurement of tensor product operators, or
collective measurements on both the system and ancilla (non-separable AAPT).
Both of these measurements are performed on the same Hilbert space $\mathcal{%
H}_{AB}$. Furthermore, it is possible to perform a generalized measurement
or POVM by going to a larger Hilbert space. In the subsequent sections, we
discuss all of these alternative strategies and argue that the non-separable
measurement schemes (whether in the same Hilbert space or in a larger one)
have hardly any practical relevance in the context of QPT, because they
require many-body interactions which are experimentally unavailable.

\subsection{Joint separable measurements}

\label{sec:JSM}

Let us assume that the initial state of the system and ancilla is $\rho
_{AB}=\sum_{ij}\rho _{ij}E_{i}^{A}\otimes E_{j}^{B}$, where $\{E_{m}^{A}\}$ (%
$\{E_{n}^{B}\}$) is the operator basis for the linear operators acting on $%
\mathcal{H}_{A}$ ($\mathcal{H}_{B}$), as defined earlier. The output state,
after applying the unknown map $\mathcal{E}$ on the principal system, is the
following:
\begin{eqnarray}
\rho _{AB}^{\prime } &=&(\mathcal{E}_{A}\otimes I_{B})(\rho _{AB})  \nonumber
\\
&=&\sum_{ij,mn}\rho _{ij}\chi _{mn}E_{m}^{A}E_{i}^{A}E_{n}^{A\dagger
}\otimes E_{j}^{B}  \nonumber \\
&=&\sum_{kj}\tilde{\alpha}_{kj}E_{k}^{A}\otimes E_{j}^{B}.
\end{eqnarray}%
In the last line, we have used $\tilde{\alpha}_{kj}=\sum_{mni}\chi _{mn}\rho
_{ij}\alpha _{k}^{m,i,n}$, where $\alpha _{k}^{m,i,n}$ is defined via $%
E_{m}^{A}E_{i}^{A}E_{n}^{A\dagger }=\sum_{k}\alpha _{k}^{m,i,n}E_{k}^{A}$,
and depends only on the choice of operator basis. From the above equation it
is clear that if we consider the basis operators as observables,\footnote{%
If $\text{dim}(\mathcal{H})=d$, then one can choose $E_{0}=\frac{1}{\sqrt{d}}%
~I$ ($d\times d$ identity matrix) and $E_{j}$ ($j=1,\ldots ,d^{2}-1$) to be
traceless Hermitian matrices.} then the parameters $\widetilde{\alpha }_{kj}$%
, which are related to the $\chi _{mn}$'s, can be obtained by joint
measurement of the observables $E_{k}^{A}\otimes E_{j}^{B}$. In fact, the
expectation values $\text{Tr}(\rho _{AB}^{\prime }E_{k}^{A\dagger }\otimes
E_{j}^{B\dagger })$, as the measurement results, are exactly the $\widetilde{%
\alpha }_{kj}$ parameters:%
\begin{equation}
\widetilde{\alpha }_{kj}=\text{Tr}(\rho _{AB}^{\prime }E_{k}^{A\dagger
}\otimes E_{j}^{B\dagger }).  \label{eq:JSM}
\end{equation}
Now, by defining
\begin{equation}
\widetilde{\chi }_{ki}=\sum_{mn}\alpha _{k}^{m,i,n}\chi _{mn},  \label{chi}
\end{equation}%
and considering that the $\alpha $ parameters are known from the choice of
operator basis, we see that by knowledge of the $\widetilde{\alpha }_{kj}$'s
we can obtain the $\chi _{mn}$ parameters through the following matrix
equation:
\begin{equation}
\widetilde{\bm{\alpha}}=\widetilde{\bm{\chi}}~\bm{\varrho},  \label{matrix}
\end{equation}%
where $\bm{\varrho}=[\rho _{ij}]$, $\widetilde{\bm{\chi}}=[\widetilde{\chi }%
_{mn}]$, and $\widetilde{\bm{\alpha}}=[\widetilde{\alpha }_{kl}]$. This
equation implies that unambiguous and unique determination of the $%
\widetilde{\bm{\chi}}$ matrix is possible iff the $\bm{\varrho}$ matrix is
invertible. After obtaining $\widetilde{\bm{\chi}}$, by using the linear
relation of Eq.~(\ref{chi}) between $\widetilde{\bm{\chi}}$ and $\bm{\chi}$
matrices, one can easily find $\bm{\chi}$ by an inversion. \footnote{%
Eq.~(\ref{chi}) can be written formally as: $\vec{\widetilde{\bm{\chi}}}=%
\vec{\bm{\chi}}\bm{A}$, where $\vec{\widetilde{\bm{\chi}}}$ ($\vec{\bm{\chi}}
$) is a row matrix obtained by arranging elements of $\widetilde{\bm{\chi}}$
($\bm{\chi}$) in some agreed order, and $\bm{A}$ is a matrix obtained by the
corresponding reordering of the $\alpha _{k}^{m,i,n}$ parameters.}

Equation~(\ref{matrix}) implies that if we were to choose $\bm{\varrho}$ as
a multiple of the $d^{2}\times d^{2}$ identity matrix $I$, then the unknown
quantum operation, $\widetilde{\bm{\chi}}$, would be directly related to the
measurement results, $\widetilde{\bm{\alpha}}$, without the need for
inversion. However, positivity of the density matrix $\rho _{AB}$ disallows
this choice. For example, in the qubit case, it can be easily seen that the
operators $\bm{\varrho}=\frac{1}{2}~I$ results in $\rho _{AB}=\frac{1}{4}%
(I_{A}\otimes I_{B}+X_{A}\otimes X_{B}+Y_{A}\otimes Y_{B}+Z_{A}\otimes
Z_{B})$, which is physically unacceptable because of its negativity.
Conversely, Eq.~(\ref{matrix}) implies that in AAPT no choice of the initial
density matrix $\rho _{AB}$ can result in a direct (inversion-free) relation
between the measurement results and elements of the unknown map.

Next, we explicitly show that the invertibility condition of $\bm{\varrho}$
in Eq.~(\ref{matrix}) is equivalent to the condition of maximal Schmidt
number in the corresponding $\rho _{AB}$. In general, an operator $Q_{AB}$
on $\mathcal{H}_{A}\otimes \mathcal{H}_{B}$ can be written as $%
Q=\sum_{jk}Q_{jk}C_{j}\otimes D_{k}$, where $\{C_{j}\}$ ($\{D_{k}\}$) is a
fixed orthonormal basis for the space of linear operators acting on $%
\mathcal{H}_{A}$ ($\mathcal{H}_{B}$). A singular value decomposition
of the matrix $\bm{Q}\equiv \lbrack Q_{jk}]$ yields $\bm{Q}=USV$,
where $U$ and $V$ are unitary matrices and $S$ is a diagonal matrix
with non-negative entries $S_{jk}=s_{k}\delta _{jk}$ ($s_{k}$ are
the singular values of the matrix $\bm{Q}$). Using this
decomposition, we have $Q=\sum_{l}s_{l}A_{l}\otimes
B_{l}$, where the operators $\{A_{l}\equiv \sum_{j}U_{jl}C_{j}\}$ and $%
\{B_{l}\equiv \sum_{k}V_{lk}D_{k}\}$ are also orthonormal bases.
This is the Schmidt decomposition of the operator $Q$
\cite{nielsen-dynamics}. In our case, $\bm{Q}$ is the matrix
$\bm{\varrho}$. We know that $\bm{\varrho}$ is invertible iff none
of its singular values is zero, i.e. $\forall l,s_{l}\neq 0$. This,
in turn, guarantees that in the Schmidt decomposition of $\rho
_{AB}$ (counterpart of $Q$) all terms are present, that is, it has
maximal Schmidt number. This confirms that the invertibility
condition---which is necessary for the applicability of input states
in AAPT---is
exactly what was already termed faithfulness above (for more detail see Ref.~%
\cite{MasoudThesis}). In fact, even separable Werner states, $\rho
_{\epsilon }=\frac{1-\epsilon }{d^{2}}I+\epsilon |\Psi ^{-}\rangle
\langle \Psi ^{-}|$ (in which
$|\Phi^-\rangle_{AB}=(|01\rangle-|10\rangle)_{AB}/\sqrt{2}$) for
$\epsilon \leqslant \frac{1}{1+d}$ \cite{braunstein-nmr}, have
maximal Schmidt number. Therefore even classical correlation between
the system and the ancilla is sufficient for AAPT.

\subsection{Mutually unbiased bases measurements}

Ancilla-assisted quantum process tomography can also be performed by using
\textquotedblleft mutually unbiased bases\textquotedblright\ (MUB)
measurements \cite{ivanovic-mub,wootters-mub}. Let us briefly review MUB,
their properties, and physical importance in the context of quantum
measurement.

Assume that $\{|a_{i}\rangle \}_{i=0}^{d-1}$ and $\{|b_{i}\rangle
\}_{i=0}^{d-1}$ are two different basis sets for the $d$-dimensional Hilbert
space $\mathcal{H}$. They are called mutually unbiased if they fulfill the
following condition:
\[
|\langle a_{i}|b_{j}\rangle |^{2}=\frac{1}{d}~\forall i,j.
\]

As an example, for $d=2$ (the case of a single qubit) it is easy to verify
that the eigenvectors of the three Pauli matrices, $X$, $Y$, $Z$, denoted
respectively by $\{|\pm \rangle \}_{X}$, $\{|\pm \rangle \}_{Y}$, and $%
\{|0\rangle ,|1\rangle \}$, constitute a set of pairwise MUB. In general,
the maximum number of MUB for an arbitrary dimensional vector space is not
yet known, however, it has been proved that it cannot be greater than $d+1$.
In addition, for $d$ being a power of prime, it has been proved that the
number of MUB is exactly $d+1$ and explicit construction algorithms are
already known \cite{som-mub,wootters-mub}.
\begin{table}[tp]
\begin{ruledtabular}
\caption[A partitioning of the 2-qubit Pauli group such that the
eigenvectors constitute a MUB.]{\small{A partitioning of the 2-qubit
Pauli group such that the eigenvectors constitute a MUB.}}
\begin{tabular}{cccc}
MUB 1& $Z^A$ & $Z^B$ & $Z^A Z^B$\\
MUB 2& $X^A$ & $X^B$ & $X^A X^B$\\
MUB 3& $Y^A$ & $Y^B$ & $Y^A Y^B$\\
MUB 4& $X^A Z^B$ & $Y^A X^B$ & $Z^A Y^B$\\
MUB 5& $X^A Y^B$ & $Y^A Z^B$ & $Z^A X^B$\\
\end{tabular}
\label{tab-mub}
\end{ruledtabular}
\vskip 1cm
\end{table}
For the case of $n$-qubit systems ($d=2^{n}$) one can show the set of $%
4^{n}-1$ Pauli operators, $\tilde{E}_{k}\equiv \otimes _{i=1}^{n}E_{\alpha
(i,k)}^{i}$, where $E_{\alpha }\in \{I,X,Y,Z\}$, can be partitioned into $%
2^{n}+1$ distinct subsets, each consisting of $2^{n}-1$ mutually commuting
observables. All the operators in each subset have a set of joint
eigenvectors. The eigenvectors of all subsets then form MUB \cite%
{lawrence-mub,Romero-mub}. Table~\ref{tab-mub} illustrates such a
MUB based partitioning of the 2-qubit Pauli operators.

The importance of MUB which is relevant to our discussion is their
application in quantum state estimation. To determine the density matrix of
a $d$-dimensional quantum system $d^{2}-1$ independent real parameters must
be determined. The most informative (sub)ensemble measurements of an
observable $\Omega $ of the system (whose spectrum is non-degenerate)
provide $d-1$ independent data points, namely the probabilities $\text{Tr}%
(\rho \pi _{i})$, where $\Omega =\sum_{i}\omega _{i}\pi _{i}$ is the
spectral decomposition of $\Omega $ with spectrum $\omega _{i}$.\footnote{%
In the non-degenerate case there are $d-1$ orthogonal projectors $\pi _{i}$
since the Hilbert space is $d$-dimensional and $\sum_{i}\pi _{i}=I$.} Thus,
to fully determine the density matrix we must measure at least $%
(d^{2}-1)/(d-1)=d+1$ different noncommuting observables. In this sense
measurement of the observables corresponding to MUB is optimal, because this
requires the smallest possible number of noncommuting measurements.
Moreover, due to the finiteness of ensembles any repeated measurement will
give rise to statistical errors. Naturally, to reduce such errors one must
increase the size of ensembles and then repeat the measurements. However, it
has been shown that a set of $d+1$ MUB measurements provides the optimal
estimation of an unknown quantum state, i.e., generates minimal statistical
error (if such MUB exist) \cite{wootters-mub}.

\begin{figure}[tp]
\includegraphics[width=7.1cm,height=2.1cm]{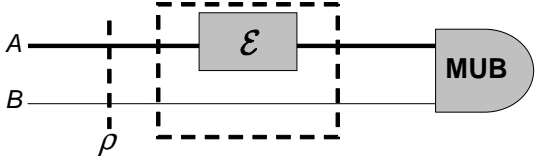}
\caption{Schematic diagram of non-separable AAPT. In this scheme the joint
separable measurements of Fig.~\protect\ref{aapt-f} have been replaced by
(collective) mutually unbiased bases measurements on the two systems.}
\label{mub-f}
\end{figure}

Now, we demonstrate that MUB measurements for state tomography yields
another version of AAPT. Let us first specialize to the single qubit case.
As noted earlier, to determine a general quantum dynamical map on a single
qubit using AAPT, one attaches an ancilla and performs quantum state
tomography at the end. In this case, the dimension of the combined Hilbert
space is $d=2^{2}$ (we assume that the dimension of the ancilla is the same
as that of the system). It follows from the general arguments above that one
can use $d+1=5$ MUB measurements to determine the final state of the
combined system (see Fig.~\ref{mub-f}). These measurements are in fact
optimal in the sense explained earlier. The first (as always, ensemble)
measurement provides four independent outcomes and each of the remaining
ones yields three independent results, which totals, as required, $%
4+(4\times 3)=16$ results.

It should be noted that even if the local state of the ancilla is known
(i.e., if we know the expectation values of $I^{A}\otimes I^{B}$, $%
I^{A}\otimes X^{B}$, $I^{A}\otimes Y^{B}$ and $I^{A}\otimes Z^{B}$ from
prior knowledge about the preparation and trace-preserving property of the
quantum map), the number of required measurements is still five \cite%
{ziman-note,mohseni-reply}. This can easily be seen from Table~\ref{tab-mub}%
. For a non trace-preserving map we need five measurements of the
(commuting) operators of the first and the second columns (the elements of
the third column are products of the operators in the first two columns). If
we know the local state of the ancilla $B$, the first three measurements of
the second column are redundant. However, since the operators in the first
column do not commute we still need to perform three (ensemble) measurements
corresponding to the first three rows. The remaining two measurements
related to the fourth and fifth rows are also necessary and they correspond
to measuring the correlations of the principal qubit and the ancilla. Thus,
the overall number of required MUB measurements in the case of
trace-preserving maps is still $5$. This argument is independent of the
basis chosen, because in any other basis, due to noncommutativity of the
Pauli operators, the measurements corresponding to the local state of the
ancilla always appear in different rows.

For the case of $n$-qubit AAPT, the dimension of the joint system-ancilla
Hilbert space is $d=2^{2n}$. In this Hilbert space four different strategies
can be devised: (i) using $16^{n}$ (separable) joint single-qubit
measurements on the $n$-qubit system and the $n$-qubit ancilla (as explained
earlier---Fig.~\ref{aapt-f}), (ii) using $5^{n}$ MUB based measurements
(tensor products of MUB based measurements of two-qubit systems), (iii)
using $d+1=4^{n}+1$ MUB based measurements on all $2n$ qubits, or (iv) using
different combinations of single-, two-, and multi-qubit measurements
including MUB based measurements (the number of measurements ranges from $%
4^{n}+1$ to $16^{n}$). In what follows, we focus on method (iii) because it
is the most economical in terms of the total number of measurements.
The main drawback of performing a MUB based measurement on all $2n$ qubits
is that it requires many-body interactions between all $2n$ qubits. From an
experimental point of view, such many-body interactions are not naturally
available. This does not mean that they cannot be simulated, but as we will
see this comes at a high resource cost. This is a strong restriction which
seriously affects the advantage of method (iii). According to our earlier
discussion, the general multi-qubit observables in a MUB based measurement
are generated from $2^{2n}+1$ noncommuting classes (or partitions) of $2n$%
-qubit operators $\{[\tilde{E}_{1}],\ldots ,[\tilde{E}_{2^{2n}+1}]\}$, where
each class $[\tilde{E}_{k}]$ contains $2^{2n}-1$ commuting observables, and $%
\tilde{E}_{k}\equiv \otimes _{i=1}^{n}E_{\alpha (i,k)}^{i}$ with $E_{\alpha
}^{i}\in \{I,X^{i},Y^{i},Z^{i}\}$ (for the case of three qubits refer to
Fig.~2 of Ref.~\cite{lawrence-mub}).

In principle, one can simulate such many-body interactions from single- and
two-qubit gates (e.g., \textsc{CNOT}). We next argue that the complexity of
such a quantum simulation scales at least as $\mathcal{O}(n^{2})$ or $%
\mathcal{O}(n^{3})$ depending, respectively, on the availability of
non-local or local MUB measurements.

All operators $\tilde{E}_{m}$ and $\tilde{E}_{n}$ that belong to the same
class $[\tilde{E}_{k}]$ commute and are composed of tensor products of
identity and/or Pauli operators. However, they cannot be simultaneously
measured locally, i.e. by using only single-qubit observables. The reason is
that each local measurement $E_{\alpha (i,m)}^{i}$ for the operator $\tilde{E%
}_{m}$ completely destroys the outcome of measuring $E_{\alpha (i,n)}^{i}$
for the other operator $\tilde{E}_{n}$, due to noncommutativity of the Pauli
operators. It is simple to see that for the non-separable measurement of an
operator such as $Z^{1}Z^{2}\ldots Z^{2n}$, we need $2n$ sequential \textsc{%
CNOT} gates. To measure a more general observable such as $\tilde{E}%
_{k}\equiv \ E_{\alpha _{1}}^{1}\ldots E_{\alpha _{2n}}^{2n}$, where $%
E_{\alpha _{i}}^{i}\in \{I,X^{i},Y^{i},Z^{i}\}$, we need $\mathcal{O}(n)$
additional single-qubit rotations to make appropriate basis changes.
Therefore, for measuring $n$ such general operators from the class $[\tilde{E%
}_{k}]$, one needs to realize $\mathcal{O}(n^{2})$ basic quantum operations.
The condition for such a construction is the possibility of addressing
arbitrary distant pairs of qubits (i.e., having access to non-local two-body
interactions). This is an important point, because in practical realizations
the spatial arrangements of the qubits or other technological reasons may
limit the interactions between distant qubits. If only nearest neighbor
gates can be implemented then pairs of qubits must be brought close to one
another (e.g., via swap gates), which incurs a cost of $\mathcal{O}(n)$
operations per pair \cite{mottonen-gates}. In this case, we need $\mathcal{O}%
(n^{3})$ quantum gates to simulate the required multi-qubit measurements. It
should also be noted that in such a simulation the scaling of execution time
and possible (operational) errors in the measurements will introduce
additional experimental complications.

\subsection{Generalized measurement}

In principle, it is also possible to perform the required quantum
state tomography at the output of AAPT by utilizing only a single
generalized measurement or POVM
\cite{d'ariano-universal,d'ariano-EuroPhys}. Suppose that we want to
determine an unknown state $\rho $ of our quantum system. In a $d$%
-dimensional Hilbert space characterization of $\rho $ requires
determination of $d^{2}-1$ independent real parameters. In order to
design a scheme for determination of $\rho$ by a single quantum
observable, we need to attach a $d^{\prime }$-dimensional ancilla
($B$) with a known initial
state $r$ to our principal system ($A$). In the scheme proposed in Ref.~\cite%
{d'ariano-universal}, one should measure one of the observables of the
combined system ($AB$), a \textquotedblleft universal quantum
observable\textquotedblright ,
\begin{equation}
\Omega =\sum_{a=1}^{dd^{\prime }}\lambda_{a}P_{a},  \label{uqo1}
\end{equation}%
where $\Omega $ is a normal operator and the spectrum $\lambda _{a}$ should
be non-degenerate such that the projections $P_{a}$ constitute a complete
set of $dd^{\prime }-1$ commuting observables. Since the projections all
commute, one can measure all of them simultaneously using a single
apparatus. Such (repeated ensemble) measurements provide us with $dd^{\prime
}-1$ probabilities $p_{a}=\text{Tr}(P_{a}\rho \otimes r)$.\footnote{%
In fact, as noted in \cite{alla}, the set of operators $\{\text{Tr}%
_{B}(rP_{a})\}$ constitutes in $\mathcal{H}_{A}$ a minimal
informationally complete POVM \cite{caves-povm}.} The dimension of
the ancilla must be greater than or equal to the dimension of the
system, $d^{\prime }\geqslant d$. If we take $\rho =\sum_{nm}\rho
_{nm}|n\rangle \langle m|$ and $r=\sum_{\alpha \beta }r_{\alpha
\beta }|\alpha \rangle \langle \beta |$, then we obtain the
following linear relation:
\begin{equation}
\rho \mapsto p_{a}=\sum_{mn}M_{mn}^{a}\rho _{nm},  \label{linearEQ}
\end{equation}
where $M_{mn}^{a}=\sum_{\alpha \beta }r_{\alpha \beta }\langle m\beta
|P_{a}|n\alpha \rangle $. When $d=d^{\prime }$ and the measured observable $%
\Omega $ couples $A$ and $B$ in a manner such that $\bm{M}_{a,mn}\equiv
M_{mn}^{a}$ is invertible ($\text{det}\bm{M}\neq 0$), a linear inversion can
reveal the unknown state $\rho $ \cite{alla}.

To be specific, we choose $\Omega $ as follows:
\begin{equation}
\Omega =\sum_{a=1}^{d^{2}}aE_{a}^{A}\otimes E_{a}^{B},  \label{uqo2}
\end{equation}%
where $\{E_{a}^{A}\}_{a=1}^{d^{2}}$ ($\{E_{a}^{B}\}_{a=1}^{d^{2}}$) is a set
of orthonormal basis operators for the space of linear operators on $%
\mathcal{H}_{A}$ ($\mathcal{H}_{B}$).\footnote{%
The operators $\{E_{a}\}$ should be normal:\ $[E_{a},E_{a}^{\dagger }]=0$,
which makes them observable in the sense defined in Ref.~\cite%
{d'ariano-universal}. In the multi-qubit case the basis operators can be
taken as tensor products of the Pauli operators.} Using the representation $%
T=\sum_{a}\mathrm{Tr}(TE_{a}^{\dag })E_{a}$ (for any operator $T$), it is
not hard to see that the ensemble average of an arbitrary operator $O$ (on $%
\mathcal{H}_{A}$) is equivalent to an ensemble average of the following
function of $\Omega $:
\[
F_{O}(\Omega )=\sum_{a}\frac{\text{Tr}(OE_{a}^{A\dagger })}{\text{Tr}%
(rE_{a}^{B})}E_{a}^{A}\otimes E_{a}^{B},
\]%
on $\rho \otimes r$, i.e.,
\begin{equation}
\langle O\rangle _{\rho }=\langle F_{O}(\Omega )\rangle _{\rho \otimes r}.
\end{equation}%
Therefore estimation of the ensemble average $\langle O\rangle _{\rho }$ of
an operator $O$ acting on the principal system $A$, can be achieved by
measuring $F_{O}(\Omega )$ on the joint $A$ and $B$ system. This allows for
the estimation of every ensemble average for the principal quantum system.

\begin{figure}[tp]
\includegraphics[width=7cm,height=2.9cm]{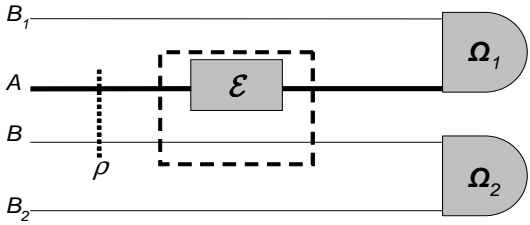}
\caption{Schematic diagram of a QPT by using POVM. Here we have used the
idea of ``universal quantum observable" \protect\cite{d'ariano-universal}.
To accomplish complete process tomography, one needs two more ancillas $%
B_{1} $ and $B_{2}$ (in addition to the one used in AAPT, $B$) and two
universal quantum observable $\Omega _{1}$ and $\Omega _{2}$.}
\label{uqo-f}
\end{figure}

The above general scheme can also be utilized for quantum process tomography
(Fig.~3 in Ref.~\cite{d'ariano-universal}). It is sufficient to consider the
AAPT scheme and attach two additional ancillas (one for the system and
another for the ancilla of the AAPT scheme), and then measure jointly two
universal observables (Fig.~\ref{uqo-f}). In this manner, to characterize
the dynamics of $n$ qubits, the number of required ancillary qubits
increases from $n$ (in AAPT) to at least $3n$. This can be easily understood
via a simple counting argument. In order to extract complete information
about a quantum dynamical map on $n$ qubits (encoded by $2^{4n}$ independent
parameters of the superoperator $\bm{\chi}$) in a single measurement, one
needs a Hilbert space of dimension at least $2^{4n}$ on which the
information can be imprinted unambiguously.

There are two major disadvantages in using such a POVM compared to all other
QPT schemes. First, the POVM scheme requires a general many-body interaction
between all $2n$ qubits that are measured through each $\Omega _{i}$. This
interaction cannot be efficiently simulated, i.e., it requires an
exponential number of elementary single- and two-qubit quantum gates.
Indeed, the above universal quantum observable scheme is very difficult to
implement in practice, because it implies measuring an observable $\Omega $
(or a commuting set $\{P_{a}\}$) which thoroughly entangles the system and
the ancilla(s). There is an alternative method to implement the above scheme
\cite{alla}, by interacting system and ancilla for a specific time duration $%
\tau $ through a known unitary operator $U$ (or known Hamiltonian $H$), and
then measuring the simplest possible non-degenerate observable $\Omega $,
namely a factorized quantity $\Omega =\omega ^{A}\otimes \omega ^{B}$ \cite%
{d'ariano-universal,alla}. However, even this method still requires
a many-body interaction (through $U$) which is difficult to prepare.
The operator $\Omega $ has maximal Schmidt number and generally
cannot be simulated using a polynomial number of elementary gates.
It is known that, in general, $\mathcal{O}(4^{N})$ elementary
single- and two-qubit gates are necessary to simulate many-body
operations acting on $N$ qubits \cite{shende} (see also Ref.
\cite{nielsen-dynamics} for different measures of complexity of a
given quantum dynamics, and Ref. \cite{jozsa-entcost} for the
concept of entanglement cost of a POVM).

\section{Direct Characterization of Quantum Dynamics}

\label{DCQD-sec}

Recently a new scheme for quantum process identification was
proposed and termed \textquotedblleft direct characterization of
quantum dynamics\textquotedblright\ (DCQD)
\cite{mohseni-dcqd1,mohseni-dcqd2}. It differs in a number of
essential aspects from SQPT\ and AAPT. In DCQD, similarly to AAPT,
the degrees of freedom of an ancilla system are utilized, but in
contrast it does not require inversion of a full $d^2\times d^2$
matrix (hence ``direct"); it requires different input states; and
uses a fixed measurement apparatus (Bell state analyzer) at the
output. The main idea in DCQD is to use certain entangled states as
inputs and to perform a simple error-detecting measurement on the
joint system-ancilla Hilbert space. A combination of these input
states and measurements give rise to a direct encoding of the
elements of the quantum map into the measurement results, which
removes the need for state tomography. More precisely, by
\textquotedblleft direct\textquotedblright\ we mean that the
measured probability distributions (on an ensemble of the setting)
are rather directly related, i.e., without the need for a complete
inversion, to the elements of $\bm{\chi}$. In essence, in DCQD the
$\bm{\chi}$ matrix elements of linear quantum maps become directly
experimentally observable. For the case of single qubit, the
measurement scheme turns out to be equivalent to a Bell-state
measurement (BSM). In DCQD the choice of input states is dictated by
whether diagonal (population) or off-diagonal (coherence) elements
of the superoperator are to be determined. Population
characterization requires maximally entangled input states, while
coherence characterization requires non-maximally entangled input
states. In the following, we review the scheme for the case of
qubits. For a generalization of the scheme to higher-dimensional
quantum systems see Ref.~\cite{mohseni-dcqd2}.

Let us consider the case of a single qubit and demonstrate how to determine
all diagonal elements of the superoperator, $\{\chi _{mm}\}$, in a single
(ensemble) measurement. We choose $\{I,X^{A},Y^{A},Z^{A}\}$ as our error
operator basis acting on the principal qubit $A$. We first maximally
entangle the two qubits $A$ (the principal system) and $B$ (the ancilla) in
a Bell-state $|\Phi ^{+}\rangle =(|00\rangle +|11\rangle )_{AB}/\sqrt{2}$
(an instance of a stabilizer code), and then subject only qubit $A$ to the
map $\mathcal{E}$.

A stabilizer code is a subspace $\mathcal{H}_{C}$ of the Hilbert space of $n$
qubits that is an eigenspace of a given Abelian subgroup $\mathcal{S}$ (the
stabilizer group) of the $n$-qubit Pauli group, with eigenvalue $+1$ \cite%
{nielsen-book,gottesman-thesis}. In other words, for every $|\Psi
_{C}\rangle \in \mathcal{H}_{C}$ and all $S_{i}\in \mathcal{S}$, we have $%
S_{i}|\Psi _{C}\rangle =|\Psi _{C}\rangle $, where the $S_{i}$'s are the
stabilizer generators and $[S_{i},S_{j}]=0$ for all $i$ and $j$. Consider
the action of an arbitrary error operator $E$ on the stabilizer code state:\
$E|\Psi _{C}\rangle $. The detection of such an error will be possible if
the error operator anticommutes with (at least one of) the stabilizer
generators: $\{S_{i},E\}=0$. To see this note that
\[
S_{i}(E|\Psi _{C}\rangle )=-E(S_{i}|\Psi _{C}\rangle )=-(E|\Psi _{C}\rangle
),
\]
i.e., $E|\Psi _{C}\rangle $ is a $-1$ eigenstate of $S_{i}$. Hence
measurement of $S_{i}$ detects the occurrence of an error or no error ($-1$
or $+1$ outcomes, respectively). Measuring all the generators of the
stabilizer then yields a list of errors (\textquotedblleft
syndrome\textquotedblright ), which allows one to determine the nature of
the errors unambiguously.

The state $|\Phi ^{+}\rangle $ is a $+1$ eigenstate of the commuting
operators $Z^{A}Z^{B}$ and $X^{A}X^{B}$, i.e., it is stabilized under the
action of these stabilizer generators. It is easy to see that any
non-trivial error operator $E_{i}\in \{I,X^{A},Y^{A},Z^{A}\}$ acting on the
state of the qubit $A$ anticommutes with at least one of the stabilizer
generators, and therefore by measuring them simultaneously we can detect the
error:
\[
\begin{array}{c}
X^{A}X^{B} \\
Z^{A}Z^{B}%
\end{array}%
(E_{i}^{A}|\Phi ^{+}\rangle )=\pm (E_{i}^{A}|\Phi ^{+}\rangle ).
\]%
Note that measuring the observables $Z^{A}Z^{B}$ and $X^{A}X^{B}$ is indeed
equivalent to a BSM, and can be represented by the four projection operators
$P_{\Psi ^{\pm }}=|\Psi ^{\pm }\rangle \left\langle \Psi ^{\pm }\right\vert $
and $P_{\Phi ^{\pm }}=|\Phi ^{\pm }\rangle \langle \Phi ^{\pm }|$, where $%
|\Phi ^{\pm }\rangle =(|00\rangle \pm |11\rangle )/\sqrt{2}$ , and $|\Psi
^{\pm }\rangle =(|01\rangle \pm |10\rangle )/\sqrt{2}$ are the Bell states.
The probabilities of obtaining the no-error outcome $I$, bit-flip error $%
X^{A}$, phase-flip error $Z^{A}$, and both phase-flip and bit-flip errors $%
Y^{A}$, on qubit $A$ can be expressed as:
\begin{equation}
p_{m}=\text{Tr}[P_{m}\mathcal{E}(\rho )]=\chi _{mm},  \label{eq:pm}
\end{equation}%
where $m=0,1,2,3$, and the projectors $P_{m}$, for $m=0,1,2,3$,
correspond to the states $\Phi ^{+}$, $\Psi ^{+}$, $\Psi ^{-}$, and
$\Phi ^{-}$, respectively. Here $\mathcal{E}(\rho)$ is a shorthand
for $(\mathcal{E}\otimes I)(\rho)$. Equation~(\ref{eq:pm}) is a
remarkable result: it shows that the diagonal elements of the
superoperator are directly obtainable from an ensemble BSM. This is
the core observation that leads to the DCQD\ scheme. In particular,
we can determine the quantum dynamical populations, $\chi _{mm}$, in
a single ensemble measurement (i.e., by simultaneously measuring the
operators $Z^{A}Z^{B}$ and $X^{A}X^{B}$) on multiple copies of the
state $|\Phi ^{+}\rangle $).

To determine the coherence elements, $\chi _{m\neq n}$, a modified strategy
is needed. As the input state we take a non-maximally entangled state: $%
|\Phi _{\alpha }^{+}\rangle =\alpha |00\rangle +\beta |11\rangle $, with $%
|\alpha |,|\beta |\notin \{0,1/\sqrt{2}\}$ and
$\text{Im}(\alpha\bar{\beta})\neq 0$. The sole stabilizer of this
state is $Z^{A}Z^{B}$. The spectral decomposition of this stabilizer is $%
Z^{A}Z^{B}=P_{+1}-P_{-1}$, where $P_{\pm 1}$ are projection operators
defined as $P_{+1}=P_{\Phi ^{+}}+P_{\Phi ^{-}}$ and $P_{-1}=P_{\Psi
^{+}}+P_{\Psi ^{-}}$. Now, it is easy to see that by measuring $Z^{A}Z^{B}$
on the output state $\mathcal{E}(\rho )$, with $\rho =|\Phi _{\alpha
}^{+}\rangle \langle \Phi _{\alpha }^{+}|$, we obtain:
\begin{equation}
\text{Tr}[P_{+1}\mathcal{E}(\rho )]=\chi _{00}+\chi _{33}+2\mathrm{Re}(\chi
_{03})\langle Z^{A}\rangle ,  \label{coherenceEQ1}
\end{equation}
and
\begin{equation}
\text{Tr}[P_{-1}\mathcal{E}(\rho )]=\chi _{11}+\chi _{22}+2\mathrm{Im}(\chi
_{12})\langle Z^{A}\rangle ,
\end{equation}
where $\langle Z^{A}\rangle =\text{Tr}(\rho Z^{A})\neq 0$ because of our
choice of a non-maximally entangled input state ($|\alpha |,|\beta |\notin
\{0,1/\sqrt{2}\}$). The experimental data, Tr$[P_{\pm 1}\mathcal{E}(\rho )]$%
, are exactly the probabilities of no bit-flip error and a bit-flip error on
qubit $A$, respectively. Since we already know the $\chi _{mm}$'s from the
population measurement, we can determine $\mathrm{Re}(\chi _{03})$ and $%
\mathrm{Im}(\chi _{12})$. After measuring $Z^{A}Z^{B}$ the system is in
either of the states $\rho _{\pm 1}=P_{\pm 1}\mathcal{E}(\rho )P_{\pm 1}/%
\text{Tr}[P_{\pm 1}\mathcal{E}(\rho )]$. Next we measure the expectation
value of a normalizer operator $N$, for example $X^{A}X^{B}$, which commutes
with the stabilizer $Z^{A}Z^{B}$.\footnote{%
A normalizer operator $N$ is a unitary operator that preserves the
stabilizer subspace but is not in $\mathcal{S}$. The normalizer group $%
\mathcal{N}$ commutes with the stabilizer group $\mathcal{S}$.} We then
obtain the measurement results
\[
\text{Tr}[N\rho _{+1}]=[(\chi _{00}-\chi _{33})\langle N\rangle +2i\mathrm{Im%
}(\chi _{03})\langle Z^{A}N\rangle ]/\text{Tr}[P_{+1}\mathcal{E}(\rho )]
\]%
and
\[
\text{Tr}[N\rho _{-1}]=[(\chi _{11}-\chi _{22})\langle N\rangle -2i\mathrm{Re%
}(\chi _{12})\langle Z^{A}N\rangle ]/\text{Tr}[P_{-1}\mathcal{E}(\rho )],
\]%
where $\langle Z^{A}\rangle $, $\langle N\rangle $, and $\langle
Z^{A}N\rangle $ are all non-zero and already known. In this manner, via a
simple linear algebraic calculation, we can extract the four independent
real parameters needed to calculate the coherence elements $\chi _{03}$ and $%
\chi _{12}$. It is easy to verify that a simultaneous measurement of the
stabilizer, $Z^{A}Z^{B}$, and the normalizer, $X^{A}X^{B}$, is again nothing
but a BSM. However, in order to construct the relevant information about the
dynamical coherence, we need to calculate the expectation values of the
Hermitian operators $P_{\Phi ^{+}}\pm P_{\Phi ^{-}}$ and $P_{\Psi ^{+}}\pm
P_{\Psi ^{-}}$.

\begin{figure}[tp]
\includegraphics[width=6cm,height=1.7cm]{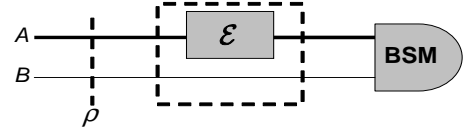}
\caption{Schematic diagram of the DCQD scheme. The system and the ancilla
are prepared in one of the input states as in Table~\protect\ref{dcqd-tab},
and after subjecting the system to the map $\mathcal{E}$, the combined
system is measured in the Bell-state basis.}
\label{bsm-f}
\end{figure}
In order to acquire complete information about the coherence elements of the
unknown dynamical map $\mathcal{E}$, we perform an appropriate change of
basis by preparing the input states $H^{A}H^{B}|\Phi _{\alpha }^{+}\rangle $
and $S^{A}S^{B}H^{A}H^{B}|\Phi _{\alpha }^{+}\rangle $, which are the
eigenvectors of the stabilizer operators $X^{A}X^{B}$ and $Y^{A}Y^{B}$. Here
$H$ and $S$ are single-qubit Hadamard and phase gates acting on the systems $%
A$ and $B.$ At the output, we measure the stabilizers and a corresponding
normalizer, e.g., $Z^{A}Z^{B}$, which are again equivalent to a standard
BSM, and can be expressed by measuring the Hermitian operators $P_{\Phi
^{+}}\pm P_{\Psi ^{+}}$ and $P_{\Phi ^{-}}\pm P_{\Psi ^{-}}$ (for the input
state $H^{A}H^{B}|\Phi _{\alpha }^{+}\rangle )$, and $P_{\Phi ^{+}}\pm
P_{\Psi ^{-}}$ and $P_{\Phi ^{-}}\pm $ $P_{\Psi ^{+}}$ (for the input state $%
S^{A}S^{B}H^{A}H^{B}|\Phi _{\alpha }^{+}\rangle $). Figure~\ref{bsm-f}
illustrates the DCQD scheme.

Overall, in DCQD we only need a single fixed measurement apparatus capable
of performing a Bell state measurement, for a complete characterization of
the dynamics. This measurement apparatus is used in four ensemble
measurements each corresponding to a different input state. Figure~\ref%
{bsm-f} and Table~\ref{dcqd-tab} summarize the preparations required for
DCQD in the single qubit case. This table implies that the required
resources in DCQD are as follows: (a) preparation of a maximally entangled
state (for population characterization), (b) preparation of three other
(non-maximally) entangled states (for coherence characterization), and (c) a
fixed Bell-state analyzer.

Our presentation of the DCQD algorithm assumes ideal (i.e., error-free)
quantum state preparation, measurement, and ancilla channels. However, these
assumptions can all be relaxed in certain situations, in particular when the
imperfections are already known. A discussion of these issues is beyond the
scope of this work and will be the subject of a future publication \cite%
{MasoudAli}.

\begingroup\squeezetable
\begin{table}[tp]
\begin{ruledtabular}
\caption{One possible set of input states and measurements for
direct characterization of quantum dynamics ($\bm{\chi}$) for a
single qubit. Here
$|\Phi^{+}_{\alpha}\rangle=\alpha|00\rangle+\beta|11\rangle$
($|\alpha|\neq 0,1/\sqrt{2}$),
$|\Phi^{+}_{\alpha}\rangle_{X(Y)}=\alpha|++\rangle_{X(Y)}+\beta|--\rangle_{X(Y)}$
($|\alpha|\neq 0,1/\sqrt{2}$ and
$\text{Im}(\alpha\bar{\beta})\neq0$) and $\{|0\rangle,|1\rangle \}$,
$\{|\pm\rangle_X \}$, $\{|\pm\rangle_Y\}$ are eigenstates of the
Pauli operators $Z$, $X$, and $Y$. The fourth column shows the BSM
measurement equivalent to stabilizer + normalizer measurements.}
\begin{tabular}{c|ccc|c}
\multicolumn{1}{c|}{input state}&\multicolumn{3}{c|}{Measurement}&\multicolumn{1}{c}{output \scriptsize{$mn$}}\\
 & Stabilizer & Normalizer & BSM& ($\chi_{mn}$)\\
 \colrule
$|\Phi^+\rangle$& $Z^AZ^B,X^AX^B$ & N/A & $P_{\Psi^{\pm}},P_{\Phi^{\pm}}$ & \tiny{00,11,22,33}\\
$|\Phi^+_{\alpha}\rangle$ & $Z^AZ^B$ & $X^AX^B$ & $P_{\Phi^{+}}\pm P_{\Phi^{-}}, P_{\Psi^{+}}\pm P_{\Psi^{-}}$ & \tiny{03,12}\\
$|\Phi^+_{\alpha}\rangle_X$ & $X^A X^B$ & $Z^A Z^B$ & $P_{\Phi^{+}}\pm P_{\Psi^{+}},P_{\Phi^{-}}\pm P_{\Psi^{-}}$ & \tiny{01,23}\\
$|\Phi^+_{\alpha}\rangle_Y$ & $Y^A Y^B$ & $Z^A Z^B$ & $P_{\Phi^{+}}\pm P_{\Psi^{-}},P_{\Phi^{-}}\pm P_{\Psi^{+}}$ & \tiny{02,13}\\
\end{tabular}\label{dcqd-tab}
\end{ruledtabular}
\end{table}
\endgroup


\section{Discussion and Resource Comparison}

\label{discussion-sec}

\begingroup\squeezetable
\begin{table*}[tp]
\begin{ruledtabular}
\caption{Required physical resources for the QPT schemes: Standard
Quantum Process Tomography (SQPT), Ancilla-Assisted Process
Tomography (AAPT) using joint separable measurements (JSM), using mutual
unbiased bases measurements (MUB), using generalized measurements
(POVM), and Direct Characterization of Quantum Dynamics (DCQD).}
\begin{tabular}{l@{\hspace{-10mm}}lcccccccc}
 Scheme & & $\text{dim}({\mathcal H})${\footnotemark[1]} & ${N}_{\text{inputs}}$ & ${N}_{\text{meas./input }}$
 {\footnotemark[2]} &   ${N}_{\text{exp.}}${\footnotemark[3]} & measurements & required interactions \\
\colrule
SQPT & & $2^n$ & $4^n$ & $4^n$ &   $16^n$   & 1-qubit & single-body \\
\multirow{3}*{AAPT} & JSM & $2^{2n}$ & 1 & $16^n$ &  $16^n$    & joint 1-qubit & single-body \\
& MUB & $2^{2n}$ & 1 &     $4^n+1$ & $4^n+1$  & MUB & many-body \\
& POVM & $2^{4n}$ & 1 & 1 &   $1$   & POVM & many-body \\
DCQD & & $2^{2n}$ & $4^n$ & 1 &   $4^n$     & BSM & single- and two- body \\
\end{tabular}
\label{tab-comp1}
\end{ruledtabular}
\footnotetext[1]{$\mathcal{H}$:
the Hilbert space of each experimental configuration}
\footnotetext[2]{total number
  of noncommuting measurements per input}
\footnotetext[3]{total number of experimental
configurations = ${N}_{\text{inputs}}~\times~{N}_{\text{meas./input }}$}
\end{table*}
\endgroup

In this section we present a discussion and comparison of the various QPT
schemes described in the previous sections, and highlight the important
features of each scheme, as illustrated in Tables~\ref{tab-comp1} and \ref%
{tab-comp2}. The goal is to provide a (physical) resource analysis and guide
for choosing the appropriate QPT\ scheme, when available resources and the
particular system of interest are taken into consideration.

\subsection{Scaling of the Required Number of Experimental Configurations with the Number of Qubits}

For characterizing a quantum dynamical map on $n$ qubits we usually perform
measurements corresponding to a tensor product of the measurements on the
individual qubits. An important example is a quantum information processing
unit with $n$ qubits. DCQD requires a total of $4^{n}$ experimental
configurations for a complete characterization of the dynamics, where the
total number of experimental configurations is defined as the number of
input states times the number of non-commuting measurements per input---see
Table~\ref{tab-comp1}. This is a quadratic advantage over SQPT and separable
AAPT, both of which require a total of $16^{n}$ experimental configurations.
However for quantum systems without controllable two-qubit operations,
implementation of the DCQD scheme is hard, here SQPT is the most efficient
scheme.

In principle, the required state tomography in AAPT could also be
realized by non-separable (global) quantum measurements. These
measurements can be performed either in the same Hilbert space, with
$4^{n}+1$ MUB measurements, or in a larger Hilbert space, with a
single generalized measurement. Both methods require many-body
interactions among $2n$ qubits, which are not naturally available.
For the AAPT scheme using MUB measurements, one can simulate the
required many-body interactions using a quantum circuit comprising
$\mathcal{O}(n^{2})$ $[\mathcal{O}(n^{3})]$ single and two-qubit
quantum elementary gates, under the assumption of available
non-local [local] two-body interactions. On the other hand, in DCQD
the only required operations are Bell-state measurements, each of
which requires one \textsc{cnot} and a Hadamard gate. This results
in a linear, $\mathcal{O}(n)$, scaling of necessary quantum
operations for realization of each experimental configuration in
DCQD---see Table~\ref{tab-comp2}.

In general, in the $2^{2n}$-dimensional Hilbert space of the $2n$ qubits of
the system and the ancilla, one could devise intermediate strategies for
AAPT using different combinations of single-, two-, and many-body
measurements. The number of experimental configurations in such methods
ranges from $4^{n}+1$ to $16^{n}$, which is always larger than that of DCQD,
which requires $4^{n}$ BSM setups. Therefore, in the given Hilbert space of $%
n$ qubits and $n$ ancillas, DCQD requires fewer experimental configurations
than all other known QPT schemes. In this sense, DCQD has an advantage over
AAPT in a Hilbert space of the same dimension. Moreover, using DCQD one can
transfer $\log _{2}2^{2n}$ bits of classical information between two
parties, Alice and Bob \cite{mohseni-dcqd1}, which is optimal according to
the Holevo bound \cite{nielsen-book}. This is a similar context to the
quantum dense coding protocol \cite{nielsen-book}. Alice can realize the
task of sending classical information to Bob by applying one of $2^{2n}$
unitary operator basis elements to the $n$ qubits in her possession and then
send them to Bob. Bob can decode the message by a single measurement on his $%
2n$ qubits using the DCQD scheme. In other words, the total number of
possible independent outcomes in each measurement in DCQD is $2^{2n}$, which
is exactly equal to the number of independent degrees of freedom for a $2n$%
-qubit system. Therefore, a maximal amount of information can be extracted
in each measurement in DCQD, which cannot be improved upon by any other
possible QPT strategy in the same Hilbert space.

For characterizing the dynamics of $n$ qubits in a single generalized (POVM)
measurement unambiguously, a Hilbert space of dimension at least $2^{4n}$ is
required. In order to implement such a POVM, one needs to realize a global
normal operator (a single universal quantum observable) acting on the joint
system-ancilla Hilbert space, of the form of Eqs.~(\ref{uqo1}) and (\ref%
{uqo2}). Such generic operators cannot be simulated in a polynomial number
of steps. It is known \cite{nielsen-book} that in general at least $\mathcal{%
O}(4^{2n})$ single- and two-qubit basic operations are needed to simulate
such general many-body operations acting on $2n$ qubits.

\subsection{Accuracy Considerations}

Due to the finiteness of the number of measurements that can be performed in
practice when estimating an ensemble average, it is evident that estimation
of an unknown quantum map through any of the QPT schemes gives rise to some
error. Such statistical errors can in principle be reduced by increasing the
size of ensembles. A relevant question in QPT discussions is then how the
accuracy of estimations in different QPT schemes depends on the ensemble
size ($N$). Finite size scaling behavior of this accuracy (or error) can
provide another practical figure-of-merit for comparison of different QPT
schemes. Here, our discussion is just tangential and very incomplete so that it just
aims at showing just a rough picture of the issue. A complete
investigation of the finite-size errors is not our goal in this paper.
Another issue that we partially address here is numerical error due to the inversion required in
some QPT\ schemes.

\subsubsection{Finite Ensemble-Size Effects}

There is a huge literature regarding analysis quantum estimation errors or quantum statistics
\cite{jezekml,hradil-ml,hradil-rehacek-prl,kosut-ml,buzek-maxent,sacchi,buzek-bayesian,blume,blume2,ziman,rohde,
helstrom-bk,holevo-bk,braunstein-caves-prl,hayashi2,gill-massar,barndroff-gill,acin-jane-vidal,matsumoto-CR,ballester-pra,
bagan,ballester-thesis,metrology,braunstein-nature,hayashi1,brody-hughston,
Qchernoffbound,nussb,deburgh,qchernoffbound2}. Our aim here is to give a very brief discussion of
estimation errors in different QPT schemes through a special example. At the end of this subsection we go a bit further
and provide a sketchy discussion of more standard figures-of-merit. However, a more complete investigation
of this subject is beyond the goals of this paper and needs a separate study per se.

In all QPT schemes measurements are performed of one or more observables $%
\{O_{k}^{(X)}\}_{k=1}^{K_{X}}$, where $X$ denotes the scheme:\ $X\in \{$%
SQPT, AAPT-JSM, AAPT-MUB, AAPT-POVM, DCQD$\}$. E.g.,
$K_{\text{AAPT-JSM}}=16^{n}$ (all operator basis for the entire Hilbert space of system and ancilla), $K_{%
\text{AAPT-POVM}}=1$ (the universal observable $\Omega $), $K_{\text{DCQD}%
}=4^{n}$ ($4$ Bell state measurements per principal qubit, one for the
superoperator population, three for the coherences---here it makes no
difference if measurements commute). For notational simplicity let us omit
the $(X)$ superscript. Each observable (given scheme $X$) has a spectral
decomposition:\ $O_{k}=\sum_{i=0}^{\nu _{k}-1}\lambda _{i}^{(k)}P_{i}^{(k)}$%
, where $\lambda _{i}^{(k)}$ are the eigenvalues and $P_{i}^{(k)}$ are
projection operators. The number $\nu _{k}$ of distinct projectors is the
number of possible measurement outcomes for a given observable $O_{k}$
(which is typically the dimension of the relevant Hilbert space). E.g., in
AAPT-POVM (where there is only a single observable), $\nu =16^{n}$, and in
DCQD $\nu _{k}=4^{n}$ for all $k$ ($n$-fold tensor product of Bell state
measurements on qubit pairs). We can also interpret $\nu _{k}$ as the
dimension of the probability space associated with a random variable $Y_{k}$
that can take values $i\in \{0,\ldots,\nu _{k}-1\}$.
\begingroup\squeezetable
\begin{table*}[tp]
\begin{ruledtabular}
\caption{Resource analysis of the QPT scheme (for the case in which
the probabilities $\{p^{(k)}_i\}$ are distributed uniformly):
Standard Quantum Process Tomography (SQPT), Ancilla-Assisted Process
Tomography using joint separable measurements (JSM), using mutual
unbiased bases measurements (MUB), using generalized measurements
(POVM), and Direct Characterization of Quantum Dynamics (DCQD).}
\begin{tabular}{l@{\hspace{-10mm}}lccccc}
 Scheme  & & ${N}_{\text{exp.}}${\footnotemark[1]} & 1-qubit gates/meas. & 2-qubit gates/meas.
 & $N_{\text{gates/meas.}}$ & $N_{\text{overall}}${\footnotemark[2]}\\
\colrule SQPT  & & $16^n$ & $\mathcal{O}(n)$ & N/A & ${\mathcal
O}(n)$
&${\mathcal O}(n 16^n)$ \\
\multirow{3}*{AAPT}& JSM  & $16^n$ & $\mathcal{O}(n)$ &
$\mathcal{O}(n)$ & ${\mathcal
O}(n)$ & ${\mathcal O}(n 16^{n})$ \\
& MUB & $4^n+1$ & $\mathcal{O}(n^2)$ &
$\mathcal{O}(n^2)~[\mathcal{O}(n^3)]$ & ${\mathcal
O}(n^2)~[{\mathcal O}(n^3)]$ & ${\mathcal O}(n^2 4^n)~[{\mathcal
O}(n^3 4^n)]$ \\
& POVM  & $1$ & ${\mathcal O}(4^{2n})$ & ${\mathcal O}(4^{2n})$
& ${\mathcal O}(4^{2n})$ & ${\mathcal O} (4^{2n})$ \\
DCQD  & & $4^n$ & $\mathcal{O}(n)$ & $\mathcal{O}(n)$ & ${\mathcal
O}(n)$ & ${\mathcal O}(n4^n)$ \\
\end{tabular}
\label{tab-comp2}
\end{ruledtabular}
\footnotetext[1]{as defined in
  Table~\ref{tab-comp1}.}
\footnotetext[2]{overall complexity = $N_{\text{exp.}}\times
N_{\text{gates/meas.}}$}
\end{table*}
\endgroup

Given an observable $O_{k}$, we must be able to unambiguously determine the
index $i$ of which projection operator (or eigenvalue) was measured. For
example, when we measure the Pauli operator $Z$, the projectors
(eigenvalues) are $P_{0}=|0\rangle \langle 0|$ ($\lambda _{0}=1$) and $%
P_{1}=|1\rangle \langle 1|$ ($\lambda _{0}=-1$), and we must have a device
(e.g., a Stern-Gerlach detector) which unambiguously reveals whether the
final state has spin up ($P_{0}$) or down ($P_{1}$). In other words, the raw
experimental outcomes are detector clicks in bins that count how many times $%
n_{i}^{(k)}$ each index $i$ has been obtained. The resulting empirical
frequencies $\{f_{i}^{(k)}\equiv n_{i}^{(k)}/N_{k}\}$, where $%
N_{k}=\sum_{i=0}^{\nu _{k}-1}n_{i}^{(k)}$, are approximations to the true
probabilities $\{p_{i}^{(k)}\}$ of detector clicks:\ $p_{i}^{(k)}=\mathrm{Tr}%
[\mathcal{E}(\rho )P_{i}^{(k)}]$. In terms of the random variable
description mentioned above, we have $\Pr (Y_{k}=i)=p_{i}^{(k)}$.

For a given observable $O_{k}$, repetition of the experiment or increase in
the sample size $N_{k}$ can reduce the error in the probability inference $%
\Delta _{i}^{(k)}\equiv |p_{i}^{(k)}-f_{i}^{(k)}|$. However, we are
in general interested in the expectation values of the observables
$O_k$, that can be obtained from the probability distribution
$\{p_{i}^{(k)}\}_{i=0}^{\nu _{k}-1}$. I.e., we would like to know
the true mean $\langle O_{k}\rangle \equiv
\mathrm{Tr}[\mathcal{E}(\rho )O_k]=\sum_{i=0}^{\nu _{k}-1}\lambda
_{i}^{(k)}p_{i}^{(k)}$, which we estimate using the empirical
frequencies to get the empirical mean $\mu _{k}\equiv
\sum_{i=0}^{\nu _{k}-1}\lambda _{i}^{(k)}f_{i}^{(k)}$. The central
limit theorem \cite{kallenberg} (or the
Chernoff inequality \cite{probability}) states that in the limit $%
N_{k}\rightarrow \infty $ the probability that the empirical mean $\mu _{k}$
is far from the expected value $\langle O_{k}\rangle $, is very small. More
precisely, defining the true standard deviation as usual as $\sigma
_{k}\equiv \sqrt{\langle O_{k}^{2}\rangle -\langle O_{k}\rangle ^{2}}$
[where $\langle O_{k}^{2}\rangle \equiv \sum_{i=0}^{\nu _{k}-1}(\lambda
_{i}^{(k)})^{2}p_{i}^{(k)}$], and letting $z_{\alpha /2}$ be the cutoff for
the upper tail of the normal distribution N$(\langle O_{k}\rangle,\sigma
_{k}^{2})$ having probability $\alpha /2$, we have asymptotically:$\ \Pr
(|\langle O_{k}\rangle -\mu _{k}|\leqslant \frac{z_{\alpha /2}}{\sqrt{N_{k}}}%
)=1-\alpha $. Here $\alpha $ represents the confidence interval. This result
allows us to compare the estimates of any two means, by equating their
confidence intervals. By replacing the true standard deviation by the
empirical one, i.e., by $\xi _{k}\equiv \sqrt{\langle \mu ^{2}\rangle
_{k}-\mu _{k}^{2}}$ where $\langle \mu ^{2}\rangle _{k}\equiv
\sum_{i=0}^{\nu _{k}-1}(\lambda _{i}^{(k)})^{2}f_{i}^{(k)}$ (an excellent
approximation in large samples), we have
\begin{equation}
\frac{\xi _{k}^{(X)}}{\sqrt{N_{k}^{(X)}}}=\frac{\xi _{k^{\prime
}}^{(X^{\prime })}}{\sqrt{N_{k^{\prime }}^{(X^{\prime })}}}  \label{eq:2}
\end{equation}
as the criterion for having a confidence interval of equal length around the
two sample means $\mu _{k}^{(X)}$\ and $\mu _{k}^{(X^{\prime })}$, i.e., to
contain the unknown true means $\langle O_{k}\rangle$\ and $\langle
O_{k^{\prime }}\rangle$\ with equal probability. Here we have reintroduced
the QPT\ label (superscript $X$) to stress that this criterion holds for the
comparison of estimates of any two means, across both $k$ and $X$. This
result shows that, assuming the standard deviations do not scale with $\nu
_{k}$, equal confidence in estimates of two expectation values of two
observables $O_{k}^{(X)}$ and $O_{k^{\prime }}^{(X^{\prime })}$ simply
requires equal sample sizes $N_{k}^{(X)}$ and $N_{k^{\prime }}^{(X^{\prime
})}$.

However, let us note that the above statistical argument is rigorous only in
the limit $N_{k}\rightarrow \infty $. That the situation is different for
finite sample sizes can be appreciated via the following examples, for which
we first recall the Chernoff inequality \cite{probability}. The version of
this inequality which is best suited to our present purpose is as follows.
Assume that an event $\gamma $ occurs with the true probability $p(\gamma )$%
. We estimate this probability by performing $N$ independent trials. The
inferred probability is then $p_{N}(\gamma )=n_{N}(\gamma )/N$, where $%
n_{N}(\gamma )$ is the number of occurrences of $\gamma $ in the trials.
Then for any $\Delta \in \lbrack 0,1]$ the Chernoff inequality is%
\begin{eqnarray}
\text{Pr}\left( |p_{N}(\gamma )-p(\gamma )|\geqslant \Delta p(\gamma )\right) \leqslant
e^{-p(\gamma )N\Delta ^{2}/3}.  \label{chernoff}
\end{eqnarray}
An immediate result of this inequality is the following. Let
\begin{equation}
N(\gamma ;\Delta ,\epsilon )\equiv \frac{3}{p(\gamma )\Delta ^{2}}\log \frac{%
1}{\epsilon }.  \label{chernoff2}
\end{equation}
Then for any $\Delta ,\epsilon \in \lbrack 0,1]$, if $N\geqslant N(\gamma
;\Delta ,\epsilon )$, then with probability greater than $1-\epsilon
$ we have $|p_{N}(\gamma )-p(\gamma )|/p(\gamma )\leqslant \Delta $.
Roughly, if we wish $p_{N}(\gamma )$ to be within an error of at
most $\Delta$ from $p(\gamma )$, this can happen with a probability
greater than $1-\epsilon $ (for some $\epsilon $) when we perform at
least $N(\gamma ;\Delta ,\epsilon )$ trials. It follows that a
highly accurate estimation ($\Delta ,\epsilon \rightarrow 0$)
requires many ($N\rightarrow \infty $) trials. In standard
statistical error analysis $\Delta $ is usually taken to be the
standard deviation $\sigma _{N} $ or at most $2\sigma _{N}$.

From Eq.~(\ref{chernoff2}) it is evident that if the probabilities
$\{p_i\}$ does not depend on the dimension of the Hilbert space
($n$) the number of repetitions to fulfill an error $\epsilon$ would
not either---this number will only have a logarithmic dependence on
the error $\epsilon$. This implies that there are cases in which the
statistics can be built up with a constant overhead in ensemble
size--up to the logarithmic dependence on the error. This can have
highly useful and efficient applications for QPT in such cases.
Nonetheless, it would be important to point out an intricate pitfall
in (incautious) too general conclusions. To this aim, here we want to
analyze a somewhat pathological example in which efficiency cannot
be concluded from Eq.~(\ref{chernoff2}). Let us assume that \emph{we
are dealing with fairly uniform probability distributions}
$\{p_{i}^{(k)}\}_{i=0}^{\nu _{k}-1}$ and compare two situations:\
tossing a coin (two possible outcomes:\ $\nu _{1}=2$) and estimating
the probability distribution of a random variable with $\nu
_{2}=10^{10}$ different possible outcomes. In the case of the coin
it is clear that after $N=10^{8}$ tosses we will have a pretty good
idea about the probabilities $p_{H}^{(1)}$ and $p_{T}^{(1)}$ of
heads vs tails. On the other hand, for the other random variable,
after $N=10^{8}$ measurements we will have not yet sampled the
entire space of possible outcomes, so will not
have been able to gather statistics representative of all probabilities $%
p_{i}^{(2)}$ (some outcomes will not have ever occurred, thus their
probabilities cannot be estimated). Consequently, we will not be able to
accurately estimate any means. However, from Eq.~(\ref{chernoff2}), with the
uniform probability distribution assumption $p_{i}^{(k)}=1/\nu _{k}$, we
obtain
\begin{eqnarray}
N_{k}\geqslant N(\Delta ,\epsilon )=3\frac{\nu _{k}}{\Delta ^{2}}\log \frac{1}{%
\epsilon }\equiv \nu _{k}C(\Delta ,\epsilon ),  \label{chernoff3}
\end{eqnarray}
In other words, for accurate estimation of means in the case of fairly
uniform probability distributions it is sufficient to have $N_{k}\geqslant C(\Delta
,\epsilon )\nu _{k}$, where the prefactor $C(\Delta ,\epsilon )$ encompasses
both the estimation error $\Delta $ and the probability $1-\epsilon $ to
achieve that error. We call condition (\ref{chernoff3}) the
``good statistics'' condition. It is
important to note that this conclusion depends strongly on the assumption of
fairly uniform probability distributions. Indeed, consider the case where
the random variable with $10^{10}$ different possible outcomes is very
strongly peaked at two values $\{i_{1},i_{2}\}$. In this case it behaves
effectively like a coin, and we do \textit{not} need $N\geqslant 10^{10}$.

We thus see that a comparison of the different QPT\ methods on the basis of
fixed mean-estimation error will depend strongly on the properties of the
underlying probability distributions $\{\{p_{i}^{(k)}\}_{i=0}^{\nu
_{k}-1}\}_{k=1}^{K_{X}}$. A thorough study of the properties of these
probability distributions as a function of QPT method $X$ is beyond the scope
of this paper. However, let us speculate on what would happen if the
assumption of fairly uniform distributions were to hold for all $k$ and $X$.
The total number of ensemble measurements becomes
\begin{equation}
N^{(X)}=\sum_{k=1}^{K_{X}}N_{k}^{(X)}\cdot N_{\text{inputs},k}^{(X)},
\label{eq:NX}
\end{equation}%
where $N_{\text{inputs},k}^{(X)}$ is the number of initial states
needed per observable $k$ for a given QPT scheme $(X)$, and
$N_{k}^{(X)}$ is found from the good statistics condition
(\ref{chernoff3}), with $\nu _{k}$ replaced by $\nu _{k}^{(X)}$. We
can read off the values of $N_{\text{inputs},k}^{(X)}$ and $K_{X}$
from the second and third columns of Table \ref{tab-comp1},
respectively. The values of $\nu _{k}^{(X)}$ are as follows. In the
case of $X$=SQPT and AAPT-JSM the observables are all
one-dimensional projectors so that $\nu _{k}=1$ $\forall k$. In the
case of $X$=AAPT-MUB there are $2n$ qubits (i.e., a
$4^{n}$-dimensional Hilbert space)\ and one performs quantum state
tomography at the output by measuring a set of noncommuting
$4^{n}+1$ observables of the MUB\ basis, where each member of the
MUB basis has a spectral resolution over $\nu _{k}=4^{n}-1$
independent projective measurements.\footnote{ One of the $4^{n}+1$
observables has $4^{n}$ independent outcomes, which gives
$(4^{n}+1)(4^{n}-1)+1=16^{n}$ outcomes, which is sufficient to fully
characterize the superoperator.} We already noted above that $\nu
_{k}^{(X)}=4^{n}$ and $16^{n}$ for $X$=DCQD, and AAPT-POVM,
respectively. We observe that, for fairly uniform distributions,
$N_{k}^{(X)}$ grows exponentially with respect to the number of
qubits $n$ for non-separable process tomography schemes, with
AAPT-MUB and AAPT-POVM at a distant disadvantage due to the inherent
depth of their quantum circuits for simulating many-body
interactions in each measurement (see the fourth column of Table
\ref{tab-comp2}). Collecting the results above, however, it follows
from Eq.~(\ref{eq:NX}) that the total number of ensemble measurements $%
N^{(X)}$ scales as $16^{n}$ in all QPT methods, to within a factor $C^{(X)}(\Delta ,\epsilon )$.

How would the number of ensemble measurements, $N_{k}^{(X)}$, change
if the distributions are sharply peaked? For separable schemes,
e.g., SQPT and AAPT-JSM, this would not result in any difference,
since we already have $\nu _{k}^{(X)}=1$. However, for non-separable
schemes this would lead to substantial reduction of measurements
since we would be dealing with effectively fixed-dimensional
probability distribution spaces, e.g., $\nu
_{k}^{(X)}=\text{const.}$, instead of an exponential function of the
number of qubits. Hence the question of the properties of the
probability distributions $\{\{p_{i}^{(k)}\}_{i=0}^{\nu
_{k}-1}\}_{k=1}^{K_{X}}$ is indeed important and will be the subject
of a future study.

\subsubsection{Discussion of Figure-of-Merit}

One of the standard approaches in quantum estimation and quantum
statistics to address estimation errors is via the Cram\'{e}r-Rao
bound (CRB) \cite{braunstein-caves-prl,
gill-massar,ballester-thesis,kosut-ml}. Following
Ref.~\cite{kosut-ml}, the CRB can be described as follows. Let us
assume that $\{\chi^{(\text{R})}_{mn}\}\in\mathbb{R}^{N^4}$ are the
true (real-valued independent) parameters of $\bm{\chi}$ that are
supposed to be estimated from a measurement data set $\mathcal{D}_X$
obtained through the scheme $X$ - we remove the superscript
$\text{R}$ in the sequel without any risk of confusion. The true
negative logarithmic likelihood function of the system generating
that true data is defined by
\begin{eqnarray*}
 \log \mathcal{L}^{(X)}=-\sum_{k=1}^{K_X}\sum_{i=0}^{\nu_k-1} n^{(k)}_i\log p^{(k)}_{i},
\end{eqnarray*}
where $n^{(k)}_i$ is the number of times the outcome $i$ is obtained
from $\ell_k$ measurements of $O_k$ (total of $\sum_k \ell
_k$ measurements) and $\langle n^{(k)}_i\rangle=p^{(k)}_i\ell_k$
(where $\langle~\rangle $ is the quantum average). If
$\widehat{\bm{\chi}}\in \mathbb{R}^{N^4}$ is an \textit{unbiased}
estimate of $\bm{\chi}$, i.e., $\langle
\widehat{\bm{\chi}}\rangle=\bm{\chi}$, the covariance of the
estimate $\text{cov}(\widehat{\bm{\chi}})=\langle (\hat{\bm{\chi}}-
\bm{\chi})(\widehat{\bm{\chi}}-\bm{\chi})^T \rangle$ satisfies the
following matrix inequality:
\begin{eqnarray}
 \left(\begin{array}{cc}\text{cov}(\widehat{\bm{\chi}}) & I \\ I & \mathcal{F}(\bm{\chi}) \end{array} \right)\geqslant 0,
\label{cr-1}
\end{eqnarray}
where $\mathcal{F}$ is the Fisher information matrix defined as
\begin{eqnarray*}
\mathcal{F}_{mn,m'n'}(\bm{\chi}) &=& \langle
\nabla_{\bm{\chi}'\bm{\chi}'}\log\mathcal{L}^{(X)}|_{\bm{\chi}}
\rangle \\
&=& \langle \frac{\partial \log p^{(k)}_i}{\partial \chi_{mn}}
\frac{\partial \log p^{(k)}_i}{\partial
\chi_{m'n'}}|_{\bm{\chi}}\rangle.
\end{eqnarray*}
Provided that $\mathcal{F}(\bm{\chi})$ is positive and invertible,
Eq.~(\ref{cr-1}) gives the following well-known form of the CRB:
\begin{eqnarray}
\text{cov}(\widehat{\bm{\chi}}) \geqslant \mathcal{F}^{-1}(\bm{\chi}).
\label{CRB}
\end{eqnarray}
Taking the trace of both sides and noting that
$\text{var}(\widehat{\bm{\chi}})=
\text{Tr}[\text{cov}(\widehat{\bm{\chi}})]$, one can also find a
scalar form for this equation. Equation~(\ref{CRB}) means that for
any unbiased estimator the error is lower-bounded by the inverse of
the Fisher information. The Fisher information matrix is indeed a
measure of information about $\bm{\chi}$ that exists in the data
$\mathcal{D}_X$. The special feature and indeed the power of this
bound is that it is independent of how the estimate is obtained, for
$\mathcal{F}$ is independent of the estimation mechanism. For the
case of single-parameter estimation, the CRB can always be achieved
\textit{asymptotically} by using maximum likelihood (ML) estimation
\cite{hradil-rehacek-prl,ballester-thesis,kallenberg}. That is, as
the amount of data increases the ML estimate approaches the true
answer with the \textit{error bars} equal to those given by the CRB.
However, for multi-parameter estimation there is, in general, no
optimal estimator that can achieve this bound. See
Ref.~\cite{ballester-thesis} and references therein for more
information about the CRB, its quantum version, and its application to
quantum state estimation. The above discussion may suggest that the
(inverse of the) Fisher information matrix can be taken as a good
figure-of-merit for a quantum estimation process. However, the very
nature of independence from estimation method means that
the Fisher information matrix is not so useful for the purpose of
comparing different QPT schemes---our goal in this
paper.

A more promising and physically-motivated approach, that justifies
using the Chernoff bound for the purpose of quantum state/process
estimation as we did earlier, has been proposed very recently, and
is called the quantum Chernoff bound (QCB)
\cite{Qchernoffbound,nussb,deburgh,qchernoffbound2}. The physical
interpretation of this quantity is as follows: assuming that we have
access to \textit{all} types of measurements---whether local or
collective---on all ensembles, the QCB measures the error in
distinguishing a state $\rho$ from another state $\widehat{\rho}$.
The probability of a wrong inference, i.e., mistaking
$\widehat{\rho}$ for $\rho$, has the asymptotic form $P_e\sim
e^{N\ln \Lambda_{\text{CB}}(\rho,\widehat{\rho})}$, where
\begin{eqnarray}
\Lambda_{\text{CB}}(\rho,\widehat{\rho})=\min_{0\leqslant \alpha\leqslant 1}\text{Tr}[\rho^{\alpha} \widehat{\rho}^{1-\alpha}],
\end{eqnarray}
is the QCB, and $0\leqslant\Lambda_{\text{CB}}\leqslant 1$. The
maximum is attained for $\rho=\widehat{\rho}$
\cite{Qchernoffbound,deburgh,qchernoffbound2}. Recently, the QCB has
been considered as a natural figure-of-merit in evaluating the
performance of different measurement scenarios for qubit tomography
\cite{deburgh}. It also has been used for quantum hypothesis testing
and distinguishability between density matrices
\cite{nussb,qchernoffbound2}. Considering the fact that a generic
$\bm{\chi}$ matrix is formally in the category of density matrices,
the application of the QCB can in principle be extended to QPT. That
is, one can in principle calculate
$\Lambda_{\text{CB}}^{(X)}(\bm{\chi},\widehat{\bm{\chi}})$ for
estimation of a quantum process $\bm{\chi}$ through any QPT scheme
$X$ and then take an average over all possible processes with a
suitable probability measure $\text{d}\mu(\bm{\chi})$
\cite{deburgh,qchernoffbound2}. The average QCB \begin{eqnarray} &
\Lambda^{(X)}_{\text{CB}}=\int\text{d}\mu(\bm{\chi})
\sum_{\widehat{\bm{\chi}}}p(\widehat{\bm{\chi}}|\bm{\chi})\Lambda_{\text{CB}}^{(X)}
(\bm{\chi},\widehat{\bm{\chi}}),
\end{eqnarray}
where $p(\widehat{\bm{\chi}}|\bm{\chi})$ is the probability of
estimating $\widehat{\bm{\chi}}$ given the true process $\bm{\chi}$,
may prove a more useful figure-of-merit. A more complete analysis,
along with possible numerics, that explicitly shows the
performance of different QPT schemes (similar to the analysis of
Ref.~\cite{deburgh}), is yet to be performed. One important point,
however, is the issue that may be caused by the assumption of
availability of all types of measurements (including collective
measurements) in this bound and whether they are important in
achieving the bound or not. This may in turn complicate usage of
this tool as a completely suitable figure-of-merit for a comparative
study of different QPT schemes. For completeness, let us mention
that a different investigation of physically good figures-of-merit
(or distance measures) for quantum operations has also been
performed in Ref.~\cite{gilchrist}.


Other characteristics of the QPT schemes may also play significant
roles in the propagation of errors in the inferred quantum map
$\mathcal{E}$. Indeed, the effect of preparation, i.e., how
different input states can affect efficiency of the estimation of
unknown maps, must be explored as well - for a recent study see
Ref.~\cite{sudarshan}. For the case of AAPT, as explained earlier,
it is already known that using maximally entangled input states is
favored, because they result in smaller experimental errors than
separable states. For DCQD an analysis of how different input states
affect performance of the estimation is underway
\cite{mohseni-rezakhani-aspuru08}. Without a full understanding of
the role of preparation, the scaleup of physical resources in
different QPT strategies for finite ensemble sizes remains elusive.
This again underlines that a promising direction is to attempt to
find a more suitable information-theoretic figure-of-merit that can
be used in a comparative finite ensemble-size analysis of the
different QPT schemes.

\subsubsection{The Role of Inversion}

It should be noted that in order to reconstruct the unknown map $\mathcal{E}$
in a QPT scheme one generally needs to perform an inversion operation which
here can be understood as Eq.~(\ref{linearEQ}). In particular, in the SQPT
and AAPT schemes an inversion on experimental data is inevitable. This
inversion may induce an ill-conditioning feature \cite%
{jezekml,boulant-lindblad}, i.e., small errors in experimental outcomes may
give rise to large errors in the estimation of $\mathcal{E}$, and can
sometimes result in non-positive\ maps. It should be stressed that quantum
dynamics obtained via the usual prescription of unitary evolution followed
by a partial trace over the bath, is always positive when the initial state
is a valid density matrix. When a positive map is applied outside of its
positivity domain it will result in non-positive density matrix. Complete
positivity results when in addition one assumes a factorized initial
system-bath state. Non-complete positivity is thus a legitimate feature of
correlated initial conditions, and non-positivity is a legitimate feature of
applying a positive map to states outside of its positivity domain \cite%
{Jordan,Carteret,Shabani-Lidar}. The problem with ill-conditioning
due to inversion is a of a different nature:\ it is a numerical
error that leads to a non-positive or non-completely-positive map.
This problem, to a large extent, can be addressed by supplementary
data analysis methods, such as ML estimation
\cite{jezekml,hradil-ml,hradil-rehacek-prl,boulant-lindblad,kosut-ml,buzek-maxent,sacchi},
Bayesian state estimation \cite{buzek-bayesian,blume,blume2}, and
other reliable regularization or reconstruction methods
\cite{ziman,rohde,d'ariano-07}. In principle, all known QPT schemes
(including DCQD) can be optimized by utilizing such statistical
error reduction techniques. Here we will not delve into the details
of such methods, as they are applicable on a similar footing to all
QPT schemes, and moreover, this issue is beyond the scope of the
present paper. However, we would like to emphasize that DCQD is
inherently more immune against such inversion-amplified errors. The
diagonal elements of a map, as discussed above, are related in DCQD
directly to measurement results. For off-diagonal elements a large
extent of directness also exists. This can easily be seen, for
example, through the determination of $\chi _{03}$ --- see Eq.~
(\ref{coherenceEQ1}) --- in which only the quantities $\chi _{00}$
and $\chi _{33}$ (already obtained from a different experimental
configuration) need to be used. That is, the formal inversion
necessary in DCQD requires only a small amount of data processing.
This, in turn implies that inversion-induced errors are amplified
less than in methods requiring a full inversion.

\subsection{Partial Characterization of Quantum Dynamics}

An important and promising advantage of DCQD is for use in partial
characterization of quantum dynamics, where one cannot afford or
does not need to carry out a full characterization of the quantum
system under study, or when one has some \textit{a priori} knowledge
about the dynamics. Using indirect QPT methods in such situations is
generally inefficient, because one must apply the whole machinery of
the scheme (including its inversion) to obtain the desired partial
information about the system. On the other hand, the DCQD scheme is
inherently applicable to the task of partial characterization of
quantum dynamics. In general, one can substantially reduce the total
number of measurements when estimating the coherence elements of the
superoperator for only specific subsets of the operator basis and/or
subsystems of interest. For example, a single ensemble measurement
is needed if one wishes to identify only the coherence elements
$\chi _{03}$ and $\chi _{12}$ of a particular qubit. In
Refs.~\cite{mohseni-dcqd1,mohseni-dcqd2,mohseni-rezakhani-aspuru,mohseni-rezakhani-aspuru08}
several examples of partial characterization have been demonstrated.
For example it was demonstrated that DCQD enables the simultaneous
determination of coarse-grained (semiclassical) physical quantities,
such as the longitudinal relaxation time $T_{1}$ and the transversal
relaxation (or dephasing) time $T_{2}$. Alternative methods for
efficient selective estimation of quantum dynamical maps has been
recently developed by utilizing random sampling \cite{Emerson07}.
The central idea of these methods is symmetrization of a quantum
channel by randomization, and then efficient estimation of gate
fidelities.  The application of such partial/selective process
estimation schemes for efficient Hamiltonian identification of open
quantum systems is important per se---besides its practical
implications---and will be presented elsewhere
\cite{mohseni-rezakhani-aspuru08}.


\section{Concluding Remarks}

In the absence of a good, reliable figure-of-merit for the performance
of QPT schemes, one cannot provide a fully fair and decisive
comparative analysis. In addition, one should also consider the
complexity of physical resources associated with noisy/imperfect
quantum state preparation and measurements. Nevertheless, in this
work we have presented a detailed \textit{resource-based} comparison of
ideal quantum process tomography schemes with respect to overall
number of experimental configurations and elementary quantum
operations. In general, SQPT is always the best approach for
complete estimation of quantum dynamical systems when controlled
two-body interactions are not either available or desirable.
However, for quantum systems with controllable single- and two-body
interactions, we have shown that the DCQD approach is more efficient
than SQPT, and all versions of AAPT, in terms of the total number of
elementary quantum operations required. For such systems, DCQD
appears attractive for near-term applications involving complete
verification of small quantum information processing units,
especially in trapped-ion and liquid-state NMR systems. For example,
the number of required experimental configurations for systems of
$3$ or $4$ physical qubits is reduced from $\sim 5\times 10^{3}$ and
$\sim 6.5\times 10^{4}$ in SQPT to $64$ and $256$ in DCQD,
respectively. Such complete characterization of quantum dynamics is
essential for verification of quantum key distribution procedures,
teleportation units (in both quantum communication and distributed
quantum computation), quantum repeaters, quantum error-correction
procedures, and more generally, in any situation in quantum physics
where a few particles interact amongst themselves and with a common
environment.

\begin{acknowledgments}
We thank J. Emerson, D. F. V. James, D. Leung, B. C. Sanders, A. M.
Steinberg, and M. Ziman for helpful discussions. This work was
supported by NSERC (to M.M.), \textit{i}CORE, MITACS, and PIMS (to
A.T.R.), and ARO W911NF-05-1-0440, NSF CCF-0523675, and NSF
CCF-0726439 to D.A.L.
\end{acknowledgments}

\end{document}